\documentclass[12pt,twoside]{article}
\begin{document}
\begin{center}
{\Large\bf Quantum Logic Network for Probabilistic Teleportation of Two-Particle State of General Form*}\\[0.6cm]

GAO Ting$^{1,2}$, WANG Zhi-xi$^1$, YAN Feng-li$^{3,4}$

{\footnotesize \sl $^1$ Department of Mathematics, Capital Normal University, Beijing 100037, China\\
$^2$ College of Mathematics and Information Science, Hebei Normal University, Shijiazhuang
050016, China\\
$^3$ Department of Physics, Hebei Normal University, Shijiazhuang 050016, China\\
$^4$ CCAST (World Laboratory), P.O. Box 8730, Beijing 100080,
China}\\[0.6cm]

\begin{minipage}{16cm}
\noindent {\footnotesize \sl  A simplification  scheme of
probabilistic teleportation of two-particle state of general form
is given. By means of the primitive operations consisting of
single-qubit gates, two-qubit controlled-not gates,
 Von Neumann measurement and classically controlled operations, we
construct an efficient quantum logical network for implementing
the new scheme of probabilistic teleportation of a two-particle
state of general form. \\
\noindent PACS: 03.65.Bz, 03.67.Hk}
\end{minipage}\\[0.6cm]
\end{center}

Quantum teleportation is one of the most striking progress  of
quantum information theory. $^1$ It will be useful in quantum
computers. $^{2,3}$ It can be used to transmit information
reliably in noisy situations where a message would otherwise be
degraded, and to transfer information from fleeting or
hard-to-control carriers to particles more suitably for permanent
storage. Moreover, it may have application in quantum cryptography
and quantum dense coding.$^{4-8}$

In 1993, Bennett et al presented the theory of quantum
teleportation which allows the transmission of an unknown qubit
state from a sender "Alice" to a spatially distant receiver "Bob"
via a quantum channel with the aid of some classical
communication.$^1$ Teleportation has drawn much attention because
its fresh notion and latent applied prospects in quantum
communication and quantum calculation.  At present, teleportation
has been generalized to many cases $^{9-15}$ and demonstrated with
the polarization photon $^{16}$ and a single coherent mode of
field $^{17}$ in the experiments. $^{18,19}$ Recently,  Shi et al
and Lu et al studied probabilistic teleportation of two-particle
entangled state via a three-particle nonmaximally entangled
state.$^{12,14}$ However, the form of two-particle entangled state
is not general form of two-particle state. In order to conquer
this
{\noindent --------------------------------------------  \\
{\footnotesize $^*$Supported by National Natural Science
Foundation of China under Grant No. 10271081 and  Hebei Natural
Science Foundation under Grant No. 101094.}}
\newpage
{\parindent=0cm limitation, Yan et al have generalized Shi's
method to the two-particle  state of general form by using partly
pure entangled four-particle state as the quantum channel. $^{15}$
In this scheme, an unknown two-particle state of general form,
whether entangled or not, can be transmitted from a sender to a
receiver with the help of partly pure entangled four-particle
state with certain probability. Unfortunately in their protocol
sixteen unitary transformations must be implemented. Obviously, it
is not favorable to  the
 experimental realization of teleportation.   More recently in terms of only single-qubit gates,
two-qubit controlled-not (CNOT) gates and measurements on single
qubits, Liu et al presented quantum logic networks for
probabilistic teleportation of a single qubit and a two-particle
entangled state, using a partially entangled pair and a
three-particle nonmaximally entangled state, respectively. $^{20}$
The quantum circuit of the probabilistic teleportation is then
constructed. Clearly the quantum logic networks for probabilistic
teleportation will be important in realizing the teleportation
scheme in the experiment. In this paper we will simplify the
scheme of Ref. [15] first, then give a quantum logic network for
probabilistic teleportation of two-particle state of general
form.}

Suppose that  an unknown two-particle
 state of general form that will be teleported is
\begin{equation}
|\phi\rangle_{12}=a|00\rangle_{12}+b|01\rangle_{12}+c|10\rangle_{12}+d|11\rangle_{12},
\end{equation}
and the  quantum channel is in the following state
\begin{equation}
|\phi\rangle_{3456}=\alpha|0000\rangle_{3456}+\beta|1001\rangle_{3456}+\gamma|0110\rangle_{3456}+\kappa|1111\rangle_{3456}.
\end{equation}
Here $a$, $b$, $c$, $d$ are arbitrary complex numbers satisfying
$|a|^2 + |b|^2 + |c|^2 + |d|^2=1$, while  $\alpha$, $\beta$,
$\gamma$, and $\kappa$ are real, and $|\alpha|$ is smaller than
others.

The particles 3 and 4 of the state $|\phi\rangle_{3456}$ and
particle pair (1,2) belong to the sender Alice, while other two
particles 5 and 6 of the state $|\phi\rangle_{3456}$ are at the
receiver Bob's side. Thus, the total state of the system is
$|\Psi\rangle_{123456}=|\phi\rangle_{12}|\phi\rangle_{3456}$. If
two Bell state measurements between particle 2 and particle 3 and
between particle 1 and particle 4 respectively are performed by
Alice, particles 5 and 6 will be collapsed into the following
unnormalized states
\begin{equation}
|\psi_0\rangle_{56}=_{14}\langle\Phi^+|_{23}\langle\Phi^+|\Psi\rangle_{123456}=\frac
{1}{2}[
a\alpha|00\rangle_{56}+b\beta|01\rangle_{56}+c\gamma|10\rangle_{56}+d\kappa|11\rangle_{56}],~~
\end{equation}
\begin{equation}
|\psi_1\rangle_{56}=_{14}\langle\Phi^-|_{23}\langle\Phi^+|\Psi\rangle_{123456}=\frac
{1}{2}[a\alpha|00\rangle_{56}+b\beta|01\rangle_{56}-c\gamma|10\rangle_{56}-d\kappa|11\rangle_{56}],~~
\end{equation}
\begin{equation}
|\psi_2\rangle_{56}=_{14}\langle\Psi^+|_{23}\langle\Phi^+|\Psi\rangle_{123456}=\frac
{1}{2}[c\alpha|00\rangle_{56}+d\beta|01\rangle_{56}+a\gamma|10\rangle_{56}+b\kappa|11\rangle_{56}],~~
\end{equation}
\begin{equation}
|\psi_3\rangle_{56}=_{14}\langle\Psi^-|_{23}\langle\Phi^+|\Psi\rangle_{123456}=\frac
{1}{2}[-c\alpha|00\rangle_{56}-d\beta|01\rangle_{56}+a\gamma|10\rangle_{56}+b\kappa|11\rangle_{56}],
\end{equation}
\begin{equation}
|\psi_4\rangle_{56}=_{14}\langle\Phi^+|_{23}\langle\Phi^-|\Psi\rangle_{123456}=\frac
{1}{2}
[a\alpha|00\rangle_{56}-b\beta|01\rangle_{56}+c\gamma|10\rangle_{56}-d\kappa|11\rangle_{56}],~~
\end{equation}
\begin{equation}
|\psi_5\rangle_{56}=_{14}\langle\Phi^-|_{23}\langle\Phi^-|\Psi\rangle_{123456}=\frac
{1}{2}[a\alpha|00\rangle_{56}-b\beta|01\rangle_{56}-c\gamma|10\rangle_{56}+d\kappa|11\rangle_{56}],~~~
\end{equation}
\begin{equation}
|\psi_6\rangle_{56}=_{14}\langle\Psi^+|_{23}\langle\Phi^-|\Psi\rangle_{123456}=\frac
{1}{2}[c\alpha|00\rangle_{56}-d\beta|01\rangle_{56}+a\gamma|10\rangle_{56}-b\kappa|11\rangle_{56}],~~~
\end{equation}
\begin{equation}
|\psi_7\rangle_{56}=_{14}\langle\Psi^-|_{23}\langle\Phi^-|\Psi\rangle_{123456}=\frac
{1}{2}[-c\alpha|00\rangle_{56}+d\beta|01\rangle_{56}+a\gamma|10\rangle_{56}-b\kappa|11\rangle_{56}],~
\end{equation}
\begin{equation}
|\psi_8\rangle_{56}=_{14}\langle\Phi^+|_{23}\langle\Psi^+|\Psi\rangle_{123456}=\frac
{1}{2}[b\alpha|00\rangle_{56}+a\beta|01\rangle_{56}+d\gamma|10\rangle_{56}+c\kappa|11\rangle_{56}],~~~
\end{equation}
\begin{equation}
|\psi_9\rangle_{56}=_{14}\langle\Phi^-|_{23}\langle\Psi^+|\Psi\rangle_{123456}=\frac
{1}{2}[b\alpha|00\rangle_{56}+a\beta|01\rangle_{56}-d\gamma|10\rangle_{56}-c\kappa|11\rangle_{56}],~~~
\end{equation}
\begin{equation}
|\psi_{10}\rangle_{56}=_{14}\langle\Psi^+|_{23}\langle\Psi^+|\Psi\rangle_{123456}=\frac
{1}{2}[d\alpha|00\rangle_{56}+c\beta|01\rangle_{56}+b\gamma|10\rangle_{56}+a\kappa|11\rangle_{56}],~~
\end{equation}
\begin{equation}
|\psi_{11}\rangle_{56}=_{14}\langle\Psi^-|_{23}\langle\Psi^+|\Psi\rangle_{123456}=\frac
{1}{2}[-d\alpha|00\rangle_{56}-c\beta|01\rangle_{56}+b\gamma|10\rangle_{56}+a\kappa|11\rangle_{56}],
\end{equation}
\begin{equation}
|\psi_{12}\rangle_{56}=_{14}\langle\Phi^+|_{23}\langle\Psi^-|\Psi\rangle_{123456}=\frac
{1}{2}[-b\alpha|00\rangle_{56}+a\beta|01\rangle_{56}-d\gamma|10\rangle_{56}+c\kappa|11\rangle_{56}],
\end{equation}
\begin{equation}
|\psi_{13}\rangle_{56}=_{14}\langle\Phi^-|_{23}\langle\Psi^-|\Psi\rangle_{123456}=\frac
{1}{2}[-b\alpha|00\rangle_{56}+a\beta|01\rangle_{56}+d\gamma|10\rangle_{56}-c\kappa|11\rangle_{56}],
\end{equation}
\begin{equation}
|\psi_{14}\rangle_{56}=_{14}\langle\Psi^+|_{23}\langle\Psi^-|\Psi\rangle_{123456}=\frac
{1}{2}[-d\alpha|00\rangle_{56}+c\beta|01\rangle_{56}-b\gamma|10\rangle_{56}+a\kappa|11\rangle_{56}],
\end{equation}
\begin{equation}
|\psi_{15}\rangle_{56}=_{14}\langle\Psi^-|_{23}\langle\Psi^-|\Psi\rangle_{123456}=\frac
{1}{2}[d\alpha|00\rangle_{56}-c\beta|01\rangle_{56}-b\gamma|10\rangle_{56}+a\kappa|11\rangle_{56}].~~
\end{equation}
Here
$$
|\Phi^\pm\rangle_{23}=\frac {1}{\sqrt
2}(|00\rangle_{23}\pm|11\rangle_{23}), |\Psi^\pm\rangle_{23}=\frac
{1}{\sqrt 2}(|01\rangle_{23}\pm|10\rangle_{23});$$
$$|\Phi^\pm\rangle_{14}=\frac {1}{\sqrt
2}(|00\rangle_{14}\pm|11\rangle_{14}), |\Psi^\pm\rangle_{14}=\frac
{1}{\sqrt 2}(|01\rangle_{14}\pm|10\rangle_{14})$$
 are  the Bell states of the particles 2 and 3, and particles 1
 and 4 respectively.

 After the  Bell state measurements, Alice informs Bob of the
 measurement outcomes through a classical communication, then Bob
 understands exactly in which one of the sixteen states in Eqs.
 (3)-(18) is located particles 5 and 6. Now an auxiliary particle $a$ with the initial state
 $|0\rangle_a$ is introduced by Bob. On the basis
 $\{|000\rangle_{56a},$$
 |001\rangle_{56a},$ $|010\rangle_{56a},$ $|011\rangle_{56a},$ $|100\rangle_{56a},$
 $|101\rangle_{56a},$ $
 |110\rangle_{56a},$ $|111\rangle_{56a}\}$, Bob will perform a collective
 unitary transformation
 \begin{equation}
 U_0=\left (
\begin{array} {cccccccc}
1&0&0&0&0&0&0&0\\
0&-\frac {\alpha}{\beta}&\sqrt {1-\frac {\alpha^2}{\beta^2}}&0&0&0&0&0\\
0&\sqrt {1-\frac {\alpha^2}{\beta^2}}&\frac {\alpha}{\beta}&0&0&0&0&0\\
0&0&0&-\frac {\alpha}{\gamma}&\sqrt {1-\frac {\alpha^2}{\gamma^2}}&0&0&0\\
0&0&0&\sqrt {1-\frac {\alpha^2}{\gamma^2}}&\frac {\alpha}{\gamma}&0&0&0\\
0&0&0&0&0&-1&0&0\\
0&0&0&0&0&0&\frac {\alpha}{\kappa}&\sqrt {1-\frac {\alpha^2}{\kappa^2}}\\
0&0&0&0&0&0&\sqrt {1-\frac {\alpha^2}{\kappa^2}}&-\frac {\alpha}{\kappa}\\
\end{array}\right )
\end{equation}
  on the  state of particles 5, 6 and $a$. After that  a
 measurement on the auxiliary qubit $a$ is performed by Bob. If the result is $|1\rangle_a$, the teleportation fails.
 However if the measurement outcome
 $|0\rangle_a$ is obtained,  the state of the particles 5 and 6 will be one of the following states
\begin{equation}
|\varphi_0\rangle_{56}=
a|00\rangle_{56}+b|01\rangle_{56}+c|10\rangle_{56}+d|11\rangle_{56},~~~~
\end{equation}
\begin{equation}
|\varphi_1\rangle_{56}=a|00\rangle_{56}+b|01\rangle_{56}-c|10\rangle_{56}-d|11\rangle_{56},~~~~
\end{equation}
\begin{equation}
|\varphi_2\rangle_{56}=c|00\rangle_{56}+d|01\rangle_{56}+a|10\rangle_{56}+b|11\rangle_{56},~~~~
\end{equation}
\begin{equation}
|\varphi_3\rangle_{56}=-c|00\rangle_{56}-d|01\rangle_{56}+a|10\rangle_{56}+b|11\rangle_{56},~~
\end{equation}
\begin{equation}
|\varphi_4\rangle_{56}=a|00\rangle_{56}-b|01\rangle_{56}+c|10\rangle_{56}-d|11\rangle_{56},~~~~
\end{equation}
\begin{equation}
|\varphi_5\rangle_{56}=a|00\rangle_{56}-b|01\rangle_{56}-c|10\rangle_{56}+d|11\rangle_{56},~~~
\end{equation}
\begin{equation}
|\varphi_6\rangle_{56}=c|00\rangle_{56}-d|01\rangle_{56}+a|10\rangle_{56}-b|11\rangle_{56},~~~
\end{equation}
\begin{equation}
|\varphi_7\rangle_{56}=-c|00\rangle_{56}+d|01\rangle_{56}+a|10\rangle_{56}-b|11\rangle_{56},~
\end{equation}
\begin{equation}
|\varphi_8\rangle_{56}=b|00\rangle_{56}+a|01\rangle_{56}+d|10\rangle_{56}+c|11\rangle_{56},~~~
\end{equation}
\begin{equation}
|\varphi_9\rangle_{56}=b|00\rangle_{56}+a|01\rangle_{56}-d|10\rangle_{56}-c|11\rangle_{56},~~~
\end{equation}
\begin{equation}
|\varphi_{10}\rangle_{56}=d|00\rangle_{56}+c|01\rangle_{56}+b|10\rangle_{56}+a|11\rangle_{56},~~
\end{equation}
\begin{equation}
|\varphi_{11}\rangle_{56}=-d|00\rangle_{56}-c|01\rangle_{56}+b|10\rangle_{56}+a|11\rangle_{56},
\end{equation}
\begin{equation}
|\varphi_{12}\rangle_{56}=-b|00\rangle_{56}+a|01\rangle_{56}-d|10\rangle_{56}+c|11\rangle_{56},
\end{equation}
\begin{equation}
|\varphi_{13}\rangle_{56}=-b|00\rangle_{56}+a|01\rangle_{56}+d|10\rangle_{56}-c|11\rangle_{56},
\end{equation}
\begin{equation}
|\varphi_{14}\rangle_{56}=-d|00\rangle_{56}+c|01\rangle_{56}-b|10\rangle_{56}+a|11\rangle_{56},
\end{equation}
\begin{equation}
|\varphi_{15}\rangle_{56}=d|00\rangle_{56}-c|01\rangle_{56}-b|10\rangle_{56}+a|11\rangle_{56},~~
\end{equation}
according to the outcomes of Bell state measurements. The teleportation can be successfully achieved with the
classical information from Alice and a corresponding unitary operation which is easy designed on the particles 5
and 6. For example we can transform
$|\varphi_{12}\rangle_{56}=-b|00\rangle_{56}+a|01\rangle_{56}-d|10\rangle_{56}+c|11\rangle_{56}$ into the  state
$a|00\rangle_{56}+b|01\rangle_{56}+c|10\rangle_{56}+d|11\rangle_{56}$ by using $\sigma_z\sigma_x$ on the particle
6, where $\sigma_x$, $\sigma_z$ are Pauli operators. In this new scheme we only use one unitary transformation
$U_0$, which can carry out the task with the same probability as $U_0$, $U_1$, ... ,$U_{15}$
 of Ref. [15]. This will make the teleportation easily realized.

Evidently the implementation of $U_0$ plays an essential  role in
the teleportation of two-particle state of general form. Barenco
et al showed that a set of gates consisting of all one-bit quantum
gates and the two-bit controlled-not (CNOT) gates is universal in
the sense that all unitary operation on arbitrarily many bits can
be expressed as compositions of these gates. They investigated a
general simulation of a three-bit controlled-u gate for an
arbitrary  one-bit unitary operation $u$ using only these basic
gates. $^{21}$ It is tedious but straightforward to prove that
$U_0$ can be expressed as
\begin{equation}
\begin{array}{l}U_0=(I\bigotimes\Lambda_2(X))C_{13}(I\bigotimes\Lambda_1(X))C_{12}(I\bigotimes\Lambda_2(X))C_{13}
(I\bigotimes X\bigotimes I)(\Lambda_1(X)\bigotimes I)(I\bigotimes
\Lambda_2(X))C_{13}\\
~~~~~~~~(I\bigotimes\Lambda_2(X))C_{13}(I\bigotimes\Lambda_1(X))C_{12}(I\bigotimes\Lambda_2(X))C_{13}(I\bigotimes
I\bigotimes X )(I\bigotimes
\Lambda_1(X))C_{23}C_{13}\\
~~~~~~~~(I\bigotimes\Lambda_2(X))C_{13}(I\bigotimes\Lambda_1(X))C_{12}(I\bigotimes\Lambda_2(X))C_{13}
(I\bigotimes X\bigotimes I)(\Lambda_1(X)\bigotimes I)\\
~~~~~~~~(I\bigotimes
\Lambda_2(X))C_{13}(I\bigotimes\Lambda_2(X))C_{13}(I\bigotimes\Lambda_1(X))C_{12}(I\bigotimes\Lambda_2(X))C_{13}(I\bigotimes
I\bigotimes X )\\
~~~~~~~~(I\bigotimes \Lambda_1(X))C_{13}C_{23}(I\bigotimes
\Lambda_1(X))C_{12}(I\bigotimes\Lambda_2(X))C_{13}(I\bigotimes\Lambda_1(X))C_{12}(I\bigotimes\Lambda_2(X))\\
~~~~~~~~C_{13}C_{12}(\Lambda_2(X)\bigotimes I)C_{23}C_{12}\Lambda
_{23}(u_1)C_{12}(\Lambda_2(X)\bigotimes
I)C_{23}C_{12}C_{13}(I\bigotimes\Lambda_2(X))C_{13}\\
~~~~~~~~(I\bigotimes\Lambda_1(X))C_{12}(I\bigotimes\Lambda_2(X))C_{13}
(I\bigotimes X\bigotimes I)(\Lambda_1(X)\bigotimes I)(I\bigotimes
\Lambda_2(X))C_{13}\\
~~~~~~~~(I\bigotimes\Lambda_2(X))C_{13}(I\bigotimes\Lambda_1(X))C_{12}(I\bigotimes\Lambda_2(X))C_{13}(I\bigotimes
I\bigotimes X )(I\bigotimes \Lambda_1(X))C_{13}\Lambda
_{13}(u_2)\\
~~~~~~~~C_{13}(I\bigotimes\Lambda_2(X))C_{13}(I\bigotimes\Lambda_1(X))C_{12}(I\bigotimes\Lambda_2(X))C_{13}
(I\bigotimes X\bigotimes I)(\Lambda_1(X)\bigotimes I)(I\bigotimes
\Lambda_2(X))\\
~~~~~~~~C_{13}(I\bigotimes\Lambda_2(X))
C_{13}(I\bigotimes\Lambda_1(X))C_{12}(I\bigotimes\Lambda_2(X))C_{13}(I\bigotimes
I\bigotimes X
)(I\bigotimes \Lambda_1(X))C_{13}\Lambda _{12}(u_3)\\
~~~~~~~~(I\bigotimes I\bigotimes
Z)(I\bigotimes\Lambda_2(X))C_{13}(I\bigotimes\Lambda_1(X))C_{12}(I\bigotimes\Lambda_2(X))C_{13}(I\bigotimes
\Lambda_1(X))C_{12}C_{23}C_{13}\\
~~~~~~~~(I\bigotimes\Lambda_2(X))C_{13}(I\bigotimes\Lambda_1(X))C_{12}(I\bigotimes\Lambda_2(X))C_{13}
(I\bigotimes X\bigotimes I)(\Lambda_1(X)\bigotimes I)(I\bigotimes
\Lambda_2(X))C_{13}\\
~~~~~~~~(I\bigotimes\Lambda_2(X))C_{13}(I\bigotimes\Lambda_1(X))C_{12}(I\bigotimes\Lambda_2(X))C_{13}(I\bigotimes
I\bigotimes X )(I\bigotimes \Lambda_1(X))C_{13}C_{23}\\
~~~~~~~~(I\bigotimes\Lambda_2(X))C_{13}(I\bigotimes\Lambda_1(X))C_{12}(I\bigotimes\Lambda_2(X))C_{13}
(I\bigotimes X\bigotimes I)(\Lambda_1(X)\bigotimes I)(I\bigotimes
\Lambda_2(X))\\
~~~~~~~~C_{13}(I\bigotimes\Lambda_2(X))C_{13}(I\bigotimes\Lambda_1(X))C_{12}(I\bigotimes\Lambda_2(X))C_{13}(I\bigotimes
I\bigotimes X )(I\bigotimes \Lambda_1(X)).
\end{array}
\end{equation}
Here $\Lambda_{ij}(u)$ are the three-bit controlled-u gates,
$i,j=1,2,3$, and $\Lambda_i(X)$ are controlled-NOT gates, $i=1,2$;
$$I=\left (
\begin{array} {cc}
1&0\\
0&1\\
\end{array}\right ); ~~~~~~
X=\left (
\begin{array} {cc}
0&1\\
1&0\\
\end{array}\right ); ~~~~~~Z=\left (
\begin{array} {cc}
1&0\\
0&-1\\
\end{array}\right );$$
$$\Lambda_1(X)=\left (
\begin{array} {cccc}
1&0&0&0\\
0&1&0&0\\
0&0&0&1\\
0&0&1&0\\
\end{array}\right ); ~~~~\Lambda_2(X)=\left (
\begin{array} {cccc}
1&0&0&0\\
0&0&0&1\\
0&0&1&0\\
0&1&0&0\\
\end{array}\right );$$
$$\Lambda_{23}(u_1)=\left (
\begin{array} {cccccccc}
1&0&0&0&0&0&0&0\\
0&1&0&0&0&0&0&0\\
0&0&1&0&0&0&0&0\\
0&0&0&-\frac {\alpha}{\beta}&0&0&0&\sqrt {1-\frac {\alpha^2}{\beta^2}}\\
0&0&0&0&1&0&0&0\\
0&0&0&0&0&1&0&0\\
0&0&0&0&0&0&1&0\\
0&0&0&-\sqrt {1-\frac {\alpha^2}{\beta^2}}&0&0&0&-\frac {\alpha}{\beta}\\
\end{array}\right );$$
$$\Lambda_{13}(u_2)=\left (
\begin{array} {cccccccc}
1&0&0&0&0&0&0&0\\
0&1&0&0&0&0&0&0\\
0&0&1&0&0&0&0&0\\
0&0&0&1&0&0&0&0\\
0&0&0&0&1&0&0&0\\
0&0&0&0&0&-\frac {\alpha}{\gamma}&0&\sqrt {1-\frac {\alpha^2}{\gamma^2}}\\
0&0&0&0&0&0&1&0\\
0&0&0&0&0&-\sqrt {1-\frac {\alpha^2}{\gamma^2}}&0&-\frac {\alpha}{\gamma}\\
\end{array}\right );$$
$$ \Lambda_{12}(u_3)=\left (
\begin{array} {cccccccc}
1&0&0&0&0&0&0&0\\
0&1&0&0&0&0&0&0\\
0&0&1&0&0&0&0&0\\
0&0&0&1&0&0&0&0\\
0&0&0&0&1&0&0&0\\
0&0&0&0&0&1&0&0\\
0&0&0&0&0&0&\frac {\alpha}{\kappa}&-\sqrt {1-\frac {\alpha^2}{\kappa^2}}\\
0&0&0&0&0&0&\sqrt {1-\frac {\alpha^2}{\kappa^2}}&\frac {\alpha}{\kappa}\\
\end{array}\right );$$
$$\Lambda_{12}(X)=C_{12}=\left (
\begin{array} {cccccccc}
1&0&0&0&0&0&0&0\\
0&1&0&0&0&0&0&0\\
0&0&1&0&0&0&0&0\\
0&0&0&1&0&0&0&0\\
0&0&0&0&1&0&0&0\\
0&0&0&0&0&1&0&0\\
0&0&0&0&0&0&0&1\\
0&0&0&0&0&0&1&0\\
\end{array}\right );$$
$$\Lambda_{23}(X)=C_{23}=\left (
\begin{array} {cccccccc}
1&0&0&0&0&0&0&0\\
0&1&0&0&0&0&0&0\\
0&0&1&0&0&0&0&0\\
0&0&0&0&0&0&0&1\\
0&0&0&0&1&0&0&0\\
0&0&0&0&0&1&0&0\\
0&0&0&0&0&0&1&0\\
0&0&0&1&0&0&0&0\\
\end{array}\right );$$
$$\Lambda_{13}(X)=C_{13}=\left (
\begin{array} {cccccccc}
1&0&0&0&0&0&0&0\\
0&1&0&0&0&0&0&0\\
0&0&1&0&0&0&0&0\\
0&0&0&1&0&0&0&0\\
0&0&0&0&1&0&0&0\\
0&0&0&0&0&0&0&1\\
0&0&0&0&0&0&1&0\\
0&0&0&0&0&1&0&0\\
\end{array}\right ).$$

Combining the results obtained by Barenco et al and  the Eq.(36)
we see that single qubit and CNOT gates can be used to implement
unitary operation $U_0$ and can  easily figure out the quantum
gate array of the unitary operation $U_0$. For the sake of saving
space we will not depict it out.

Based on the simplification scheme and  the decomposition of
unitary operator $U_0$ we can constructed a quantum network for
probabilistic teleportation of an unknown two-particle state of
general form. It is shown in Fig. 1.

 Here $R$, $R'$, $R''$ are single-qubit rotation transformations, and $H$ is a
 Hadamard gate defined as $H=\frac {1}{\sqrt 2}\left (\begin{array}{cc}
 1&1\\
 1&-1\\
 \end{array}\right)$. We carefully choose $R$, $R'$, and $R''$ to
 make the state of the particles 3, 4, 5, and 6  to be the quantum channel $|\phi\rangle_{3456}=
 \alpha|0000\rangle_{3456}+\beta|1001\rangle_{3456}+\gamma|0110\rangle_{3456}+\kappa|1111\rangle_{3456}$ by the first dash line.
 If the measurement result on the auxiliary
 qubit at the output state is $|0\rangle_a$, the teleportation is
 successful with the final state of the particle 5 and 6 being
 reconstructed as the initial state to be teleported. From  Fig.1,
 we see that the probabilistic teleportation can be divided into
 two steps. The first step is a purification process in which Bob
 can concentrate the state $\alpha|0000\rangle_{3456}+\beta|0110\rangle_{3456}+\gamma|1001\rangle_{3456}+\kappa|1111\rangle_{3456}
$ into a maximally entangled state by introducing an auxiliary
particle $a$ and performing a collective unitary operation $U_0$.
The second step is the standard teleportation using the maximal
entanglement as the quantum channel.

Note that a controlled unitary operation acting on any number of
qubits followed by the measurement of the control qubit can be
replaced by the measurement of the control qubit preceding  the
controlled operation. $^{20}$ Therefore, Fig.1 can be re-expressed
as Fig.2.

In Fig.2, the controlled operation can be realized locally by Bob
depending on the results of the four measurements Alice performed
on her own qubit. If and only if the outcome of Alice's
measurement is 1, Bob can execute the controlled-$R_x$ or
controlled-$R_z$, where $R_x=X$, $R_z=HR_xH$. If the measurement
result of the auxiliary particle is $|1\rangle_a$, the
teleportation fails. If the result is $|0\rangle_a$, the final
state of particles 5 and 6 at Bob's side will be collapsed into
$|\phi\rangle_{12}=a|00\rangle_{12}+b|01\rangle_{12}+c|10\rangle_{12}+d|11\rangle_{12}$,
which is the desired state. That is to say, if the measurement
outcome on the state of the auxiliary qubit is $|0\rangle_a$,
perfect teleportation is accomplished.

In summary, a simplification scheme for probabilistic teleporting
an unknown two-particle state of general form is given. By means
of the primitive operations consisting of single-qubit gates,
two-qubit controlled-not gates,  Von Neumann measurement and
classically controlled operations, we construct an efficient
quantum logical network for implementing the new scheme of
probabilistic teleportation of a two-particle state of general
form. Both the product state and entangled state of two-particles
can be transmitted from the sender to receiver by the partly pure
entangled four-particle state with certain probability. We hope
that this scheme will be realized by experiment.

\vspace {0.6cm}

{\parindent=0cm \bf REFERENCES } \footnotesize
\begin{tabbing}
xxxxx\=\kill
~$^1$ C. H. Bennett  et al, Phys. Rev. Lett.   70 (1993)  1895.\\
~$^2$ J. I. Cirac and P. Zoller, Phys. Rev. Lett. 74 (1995) 4091.\\
~$^3$ A. Barenco et al, Phys. Rev. Lett. 74 (1995)4083.\\
~$^4$ A. K. Ekert,  Phys. Rev. Lett.  67  (1991) 661.\\
~$^5$ C. H. Bennett,  Phys. Rev. Lett.  68  (1992) 3121.\\
~$^6$ Y. Zhang   et al, Chin. Phys. Lett. 15 (1998) 238.\\
~$^7$ B. S. Shi  and G. C. Guo,  Chin. Phys. Lett.  14 (1997) 521.\\
~$^8$ C. H. Bennett  and  S. J. Wiesner,  Phys. Rev. Lett.  69 (1992) 2881.\\
~$^9$ M. Ikram, S. Y. Zhu and M. S.  Zubairy, Phys. Rev. A62 (2000) 022307. \\
$^{10}$ W. L. Li, C. F. Li and G. C. Guo, Phys. Rev. A61 (2000) 034301.\\
$^{11}$ V. N. Gorbachev  and A. I. Trubilko, J. Exp. Theor. Phys.  91 (2000) 894.\\
$^{12}$ H. Lu and  G. C. Guo, Phys. Lett.   A276  (2000) 209.\\
$^{13}$ B. Zeng, X. S. Liu, Y. S. Li and G. L. Long, Commun. Theor. Phys. 38 (2002) 537.\\
$^{14}$ B. S. Shi  et al,  Phys. Lett.   A268 (2000) 161.\\
$^{15}$ F. L. Yan, H. G. Tan and L. G. Yang, Commun. Theor. Phys. 37 (2002) 649.\\
$^{16}$ D. Bouwmeester  et al,  Nature  390 (1997) 575.\\
$^{17}$ A. Furusawa  et al,  Science  282 (1998) 706.\\
$^{18}$ D. Boschi  et al, Phys. Rev. Lett.   80 (1998)  1121.\\
$^{19}$ M. A. Nielsen   et al, Nature    396 (1998)  52.\\
$^{20}$ J. M. Liu, Y. S. Zhang and G. C. Guo, Chinese Physics 12 (2003) 251.\\
$^{21}$ A. Barenco et al, Phys. Rev. A 52 (1995) 3457.\\
\end{tabbing}

\end{document}